\documentclass[twocolumn]{aastex6}
\usepackage{epstopdf}

\begin{document}


\title{The pre-penumbral magnetic canopy in the solar atmosphere}


\author{David MacTaggart}
\affil{School of Mathematics and Statisitcs 
University of Glasgow, Glasgow 
G12 8QW, Scotland, UK}

\author{Salvo L. Guglielmino and Francesca Zuccarello}
\affil{Dipartimento di Fisica e Astronomia - Sezione
Astrofisica, Universit\`{a} di Catania,
            via S. Sofia 78, 95123 Catania, Italy}


\begin{abstract}

Penumbrae are the manifestation of magnetoconvection in highly inclined (to the vertical direction) magnetic field. The penumbra of a sunspot tends to form, initially, along the arc of the umbra antipodal to the main region of flux emergence.  The question of how highly inclined magnetic field can concentrate along the antipodal curves of umbrae, at least initially, remains to be answered. Previous observational studies have suggested the existence of some form of overlying magnetic canopy which acts as the progenitor for penumbrae. We propose that such overlying magnetic canopies are a consequence of how the magnetic field emerges into the atmosphere and are, therefore, part of the emerging region. We show, through simulations of twisted flux tube emergence, that canopies of highly inclined magnetic field form preferentially at the required locations above the photosphere.

\end{abstract}

\keywords{magnetohydrodynamics (MHD)  --- 
magnetic fields  --- sunspots}



\section{Introduction} \label{sec:intro}
\begin{figure*}
	\centering
{(a)}
\includegraphics[trim=35 190 100 265, clip, scale=.325]{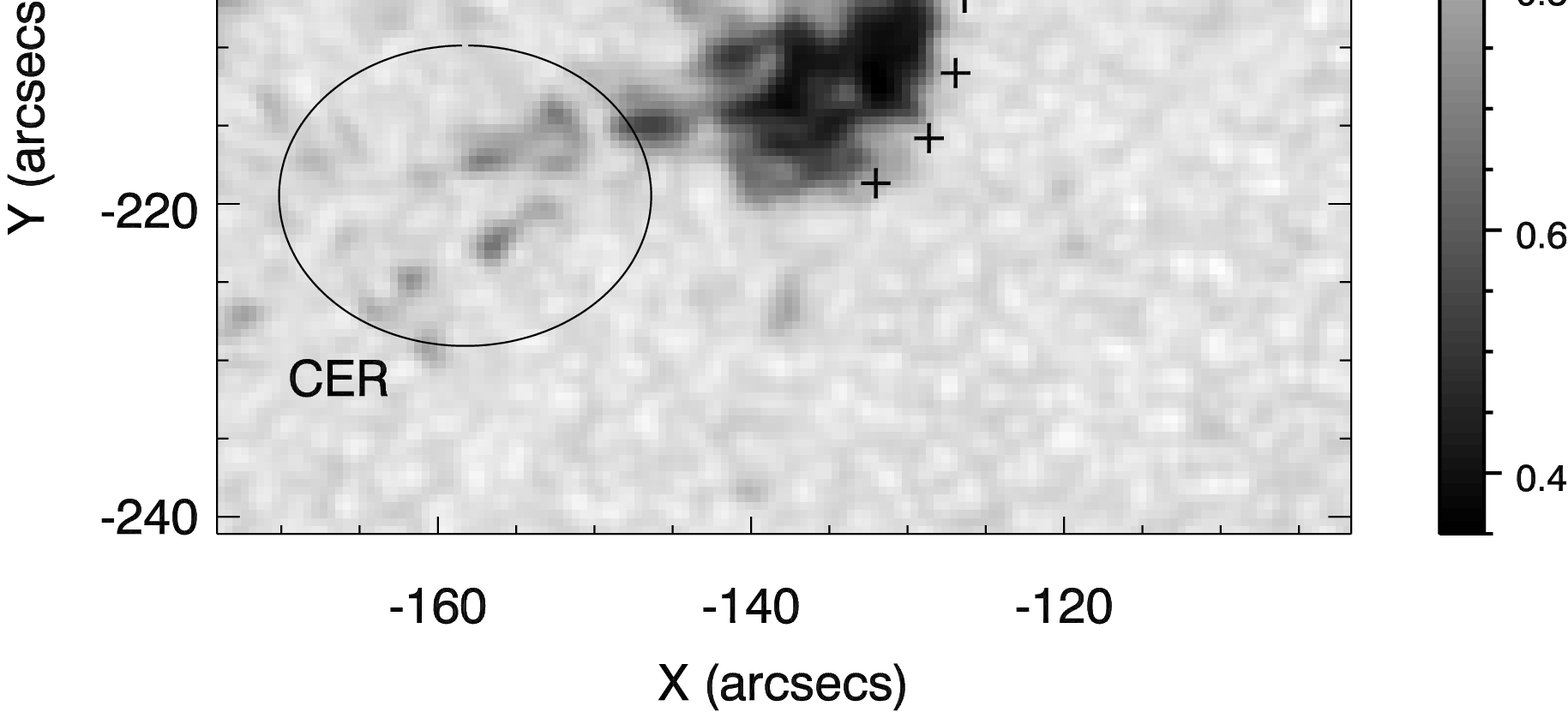}
{(b)}
\includegraphics[trim=65 190 100 265, clip, scale=.325]{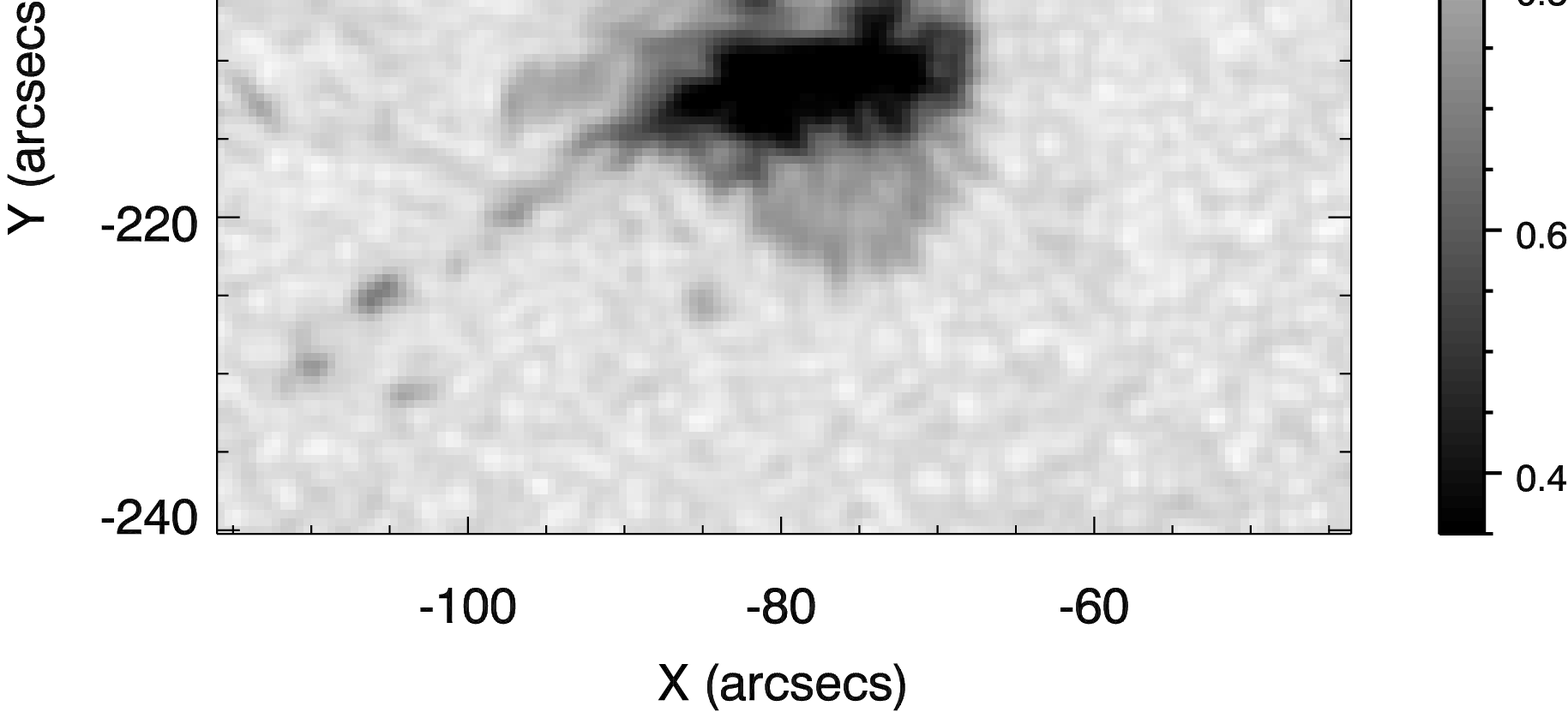}
{(c)}
\includegraphics[trim=65 190  20 265, clip, scale=.325]{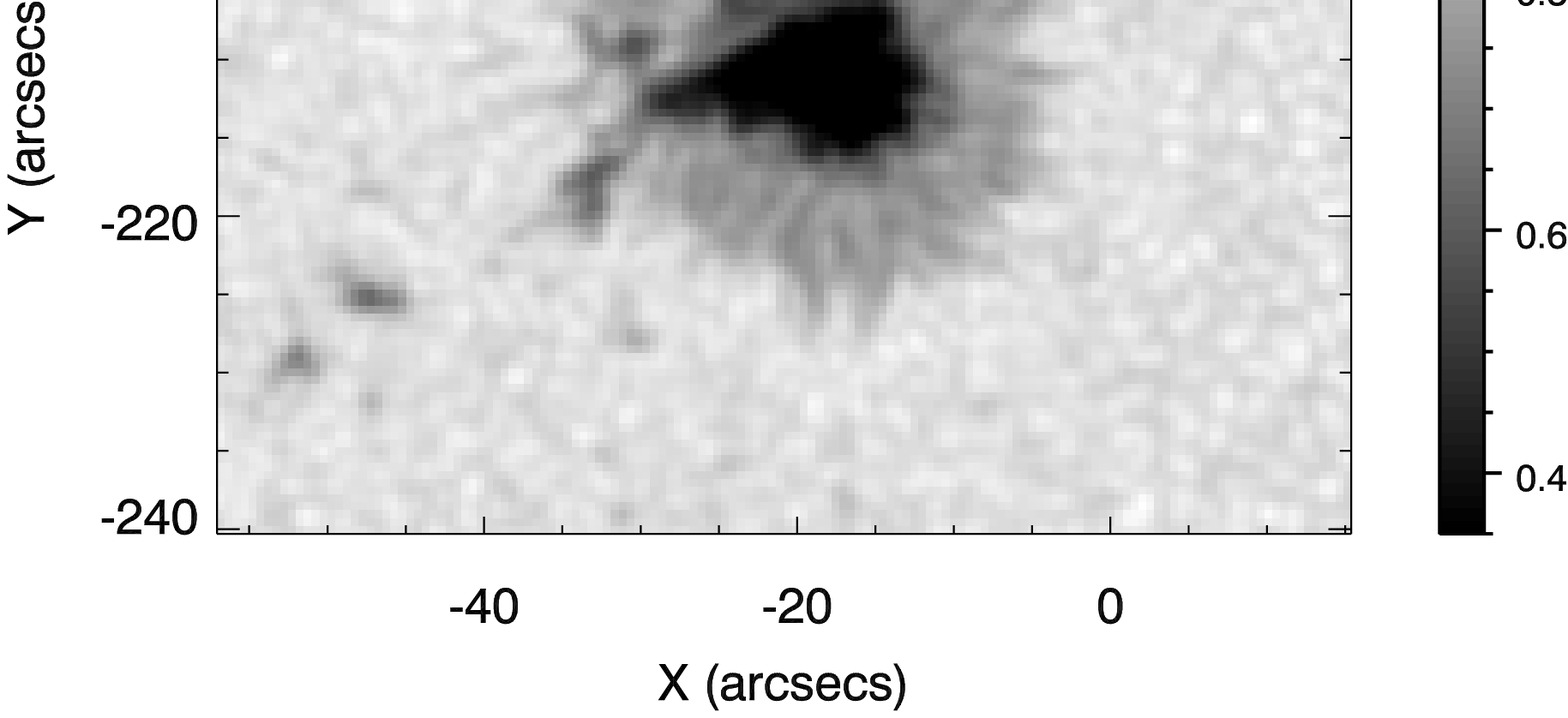}
\caption{Different stages of penumbra formation. (a) shows the pre-penumbral spot. The \emph{central emerging region}, CER, and \emph{antipodal curve}, AC, are highlighted. (b) The penumbra forms along the AC. (c) The penumbra now occupies the entire AC and is not developed only at a small location near the CER. This penumbra formation is also analyzed by \cite{romano13} and \cite{murabito16}.}
\label{obs}
\end{figure*}

A sunspot represents a strong concentration of magnetic field in the photosphere. Although a sunspot exhibits much fine-scale structure, it can be characterized by two regions with substantially different inclinations of the magnetic field. The central region, the \emph{umbra}, contains predominantly vertical field, i.e. normal to the photosphere. Surrounding the umbra is the \emph{penumbra}, where the field is much more inclined to the vertical direction. Since the umbra and penumbra sit in a convecting plasma, magnetoconvection ensues and produces much fine-scale structure \citep{thomas08}. The different dynamics of the umbra and penumbra depend on the magnetic field inclination \citep{rempel11,rempel12}. Although the `horn' geometry of a sunspot magnetic field has been known for a long time, exactly how it forms remains to be answered. Observations show that particular sections of penumbrae form first. These are typically located on the antipodal, with respect to the emerging region, arcs of the umbrae. The phenomenon has been reported in many observational studies \citep[e.g.][]{schlichenmaier10,rezaei12,shimizu12,romano13}. Figure 1 shows a sunspot at different times in the evolution of its penumbra. For our purposes we shall define two distinct spatial regions that are highlighted in Figure \ref{obs} (a). The \emph{antipodal curve}, AC, is the region where the penumbra first forms and is indicated by a border of crosses. The \emph{central emergence region}, CER, is the main emerging region between the two main active region sunspots and is indicated by an ellipse. Figure \ref{obs} (a) displays a spot before its penumbra has formed. Later, in Figure \ref{obs} (b), the penumbra grows along parts of the AC. In Figure \ref{obs} (c), the penumbra has now engulfed the AC and is fully developed except at a small location near the CER.

As a penumbra represents a region of inclined magnetic field, how is it that such field collects initially in preferential locations along the AC, as shown in Figure 1? Recent observational studies have suggested that an \emph{overlying magnetic canopy} must exist as a prelude to penumbra formation \citep{shimizu12, romano13}. \cite{shimizu12} go as far to state that ``the magnetic field structure in the chromosphere needs to be considered in the formation process of the penumbrae". There are two possible origins for an overlying magnetic canopy. The first is that it existed in the atmosphere before the emergence of the active region. The second is that the canopy is somehow connected to the emerging region. Since the first option would require the background atmosphere to combine many imponderables favourably (e.g. field inclination, direction, location, etc.) we shall focus on the second option.

 In this Letter we propose that penumbra formation is a simple consequence of how the emerging magnetic field expands into the atmosphere. We argue this through analyzing the magnetic field structure of emerged flux tubes. The rest of the Letter is outlined as follows: the model is presented, outlining the equations and modelling assumptions; the magnetic field inclination is investigated in relation to its position relative to sunspots; a discussion of the results concludes the Letter.  

\section{Model description} \label{sec:model}
In this Letter we are not concerned with producing the fine-structure of sunspot dynamics but the large-scale distribution of magnetic field inclination in an emerging region. To investigate this property we present simulations of magnetic flux emergence. The compressible and resistive magnetohydrodynamic (MHD) equations are solved using a Lagrangian remap scheme \citep{arber01}. In dimensionless form, the equations are 
\begin{equation}\label{mass_con}
{\dot{\rho}} = -\rho\nabla\cdot\mathbf{u},
\end{equation}
\begin{equation}\label{mom_con}
{\dot{\mathbf{u}}} = -\frac{1}{\rho}\nabla p + \frac{1}{\rho}(\nabla\times\mathbf{B})\times\mathbf{B}+\mathbf{g}+\frac{1}{\rho}\nabla\cdot\mathbf{T}_V,
\end{equation}
\begin{equation}\label{induction}
{\dot{\mathbf{B}}} = (\mathbf{B}\cdot\nabla)\mathbf{u} - {(\nabla\cdot\mathbf{u})\mathbf{B}} +\eta\nabla^2\mathbf{B},
\end{equation}
\begin{equation}\label{energy_con}
{\dot\varepsilon} = -\frac{p}{\rho}\nabla\cdot\mathbf{u} + \frac{1}{\rho}\eta j^2+ \frac{1}{\rho}\mathbf{T}_V:{\nabla\mathbf{u}},
\end{equation}
\begin{equation}\label{divB}
\nabla\cdot\mathbf{B} = 0,
\end{equation}
with specific energy density
\begin{equation}\label{energy_den}
\varepsilon = \frac{p}{(\gamma-1)\rho}.
\end{equation}
The over-dot represents the material derivative and the double-dot represents the double contraction of a second order Cartesian tensor. The basic variables are the density $\rho$, the pressure $p$, the magnetic field $\mathbf{B}$ and the velocity $\mathbf{u}$. $j$ is the magnitude of current density, $\mathbf{g}$ is gravity and $\gamma (=5/3)$ is the ratio of specific heats. The nondimensionalization follows that of other works \citep[e.g.][]{murray06,dmac15} with (photospheric) values for the pressure $p_{\rm ph}=1.4\times10^4$~Pa; density $\rho_{\rm ph}=3\times10^{-4}$~kg~m$^{-3}$; scale height $H_{\rm ph}=170$~km; magnetic field $B_{\rm ph}=1.3\times10^3$~G; speed $u_{\rm ph}=6.8$~km~s$^{-1}$; time $t_{\rm ph}=25$~s and temperature $T_{\rm ph}=5.6\times10^3$~K. A uniform resistivity is used, $\eta=0.001$. The viscosity tensor is given by
\begin{equation}\label{visc_ten}
\mathbf{T}_V = \mu\left(\nabla\mathbf{u}+\nabla\mathbf{u}^{\rm T}-\frac{2}{3}\mathbf{I}\nabla\cdot\mathbf{u}\right),
\end{equation}
where $\mu=0.0001$ and $\mathbf{I}$ is the identity tensor.

The idealized initial equilibrium atmosphere is given by prescribing the temperature profile
\begin{equation}\label{initial_temp}
{T(z) = \left\{\begin{array}{cc}
1-\frac{\gamma-1}{\gamma}z, & z < z_{\rm ph}, \\
1, & z_{\rm ph} \le z \le z_{\rm tr},  \\
T_{\rm cor}^{[(z-z_{\rm tr})/(z_{\rm tr}-z_{\rm ph})]}, & z_{\rm tr} < z < z_{\rm cor},  \\
T_{\rm cor}, & z \ge z_{\rm cor},
\end{array}\right.}
\end{equation}
where $T_{\rm cor} = 150$ is the initial coronal temperature, $z_{\rm ph}$ is the base of the photosphere, $z_{\rm tr}=z_{\rm ph}+10$ is the base of the transition region and $z_{\rm cor}=z_{\rm ph}+20$ is the base of the corona. In this paper, $z_{\rm ph} =$ 0. The solar interior is defined by $z<z_{\rm ph}$ and is taken, for simplicity, to be convectively stable \citep{hood12}. The other state variables, pressure and density, are found by solving the hydrostatic equation in conjunction with the ideal equation of state
\begin{equation}\label{EOS}
\frac{{\rm d}p}{{\rm d}z} = -\rho g, \quad p = \rho T.
\end{equation}
The domain size is $(x,y,z)\in[-110,110]\times[-110,110]\times[-30,80]$. The resolution is 312$^3$. 
The form of magnetic flux tube that is placed in the solar interior is similar to other studies \citep[e.g.][]{galsgaard05,murray06} and has the (cylindrical) components
\begin{equation}\label{mag_field}
B_y=B_0\exp(-r^2/R^2), \quad B_{\theta}=\alpha rB_y, \quad B_r=0,
\end{equation}
where $r^2=x^2+(z-z_0)^2$, $z_0$ is the initial height of the tube axis, $R$ is the tube radius, $B_0$ is the initial axial field strength and $\alpha$ is the twist. In this Letter we choose the values $R=3.5$ and $z_0=-20$, which are typical for flux emergence studies \citep{hood12}. We vary the other parameters in order to assess their influence on the inclination of the emerged field. The flux tube is perturbed, in order to initiate its rise, with a density deficit proportional to $\exp(-y^2/\lambda^2)$. In the following simulations we take $\lambda=15$. 

The sizes of the regions that we consider in this Letter are smaller than typical active regions, which have lengths of $O(100)$ Mm across. The regions we are modelling here have lengths of $O(30)$ Mm and are more comparable to large ephemeral regions \citep[e.g.][]{guglielmino10}. This is a modelling choice in order to be able to resolve different regions of the atmosphere. Scaling up to full active region size would result in the photoshere/chromosphere region shrinking to one or two grid points. The size of the modelled region will not have a strong effect on the results that we will present. We shall return to this point in the Discussion.

\section{Simulations}\label{sec:simulations}
To investigate how the magnetic field inclination is distributed after emergence into the atmosphere, we consider three numerical experiments with different values of the axial field strength and the twist. These are $E_1$: $B_0=6$, $\alpha_0=0.3$; $E_2$: $B_0=6$, $\alpha_0=0.2$ and $E_3$: $B_0=8$, $\alpha_0=0.3$. 
\subsection{General features}
\begin{figure}
\fig{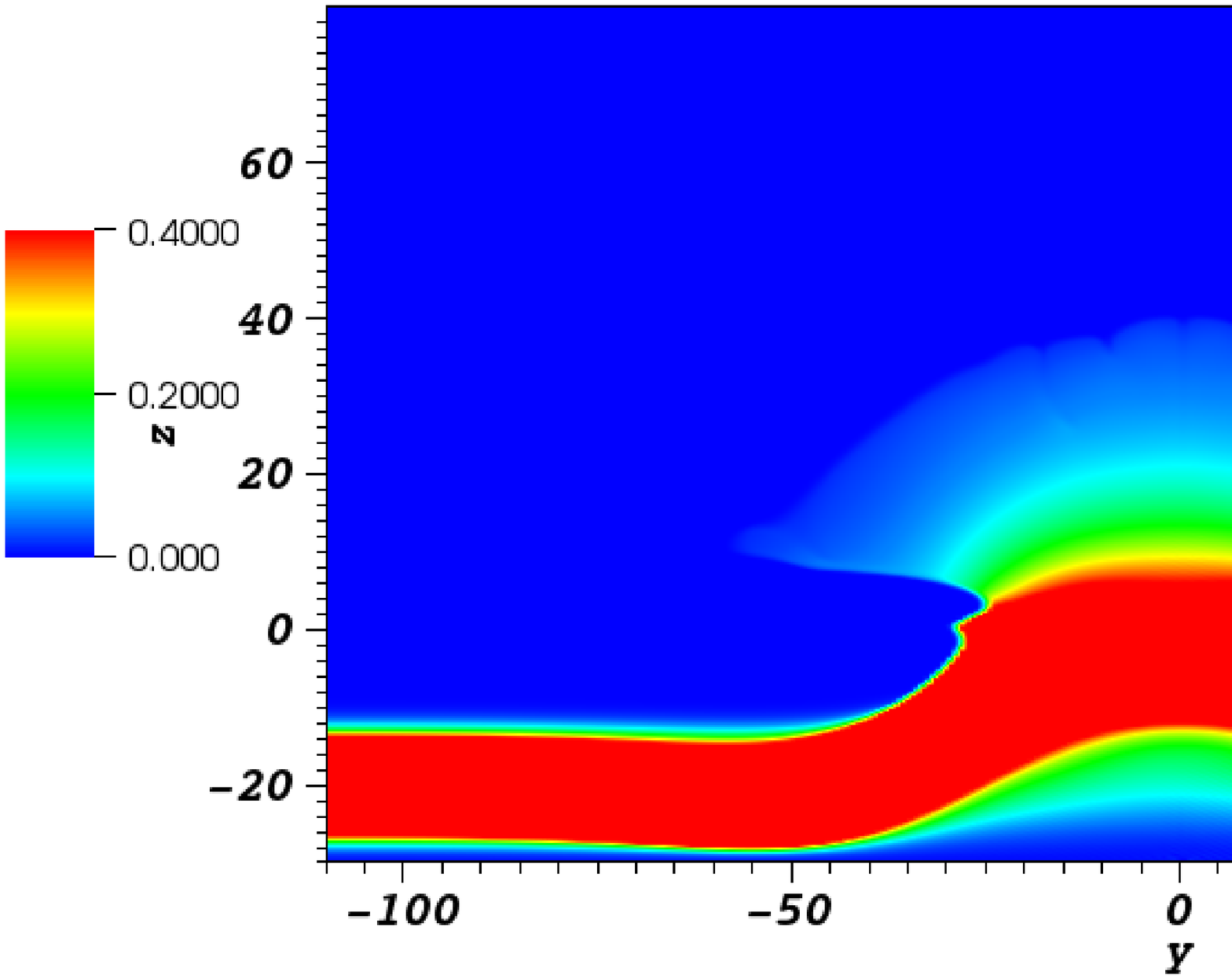}{0.5\textwidth}{(a)}\begin{center}
\fig{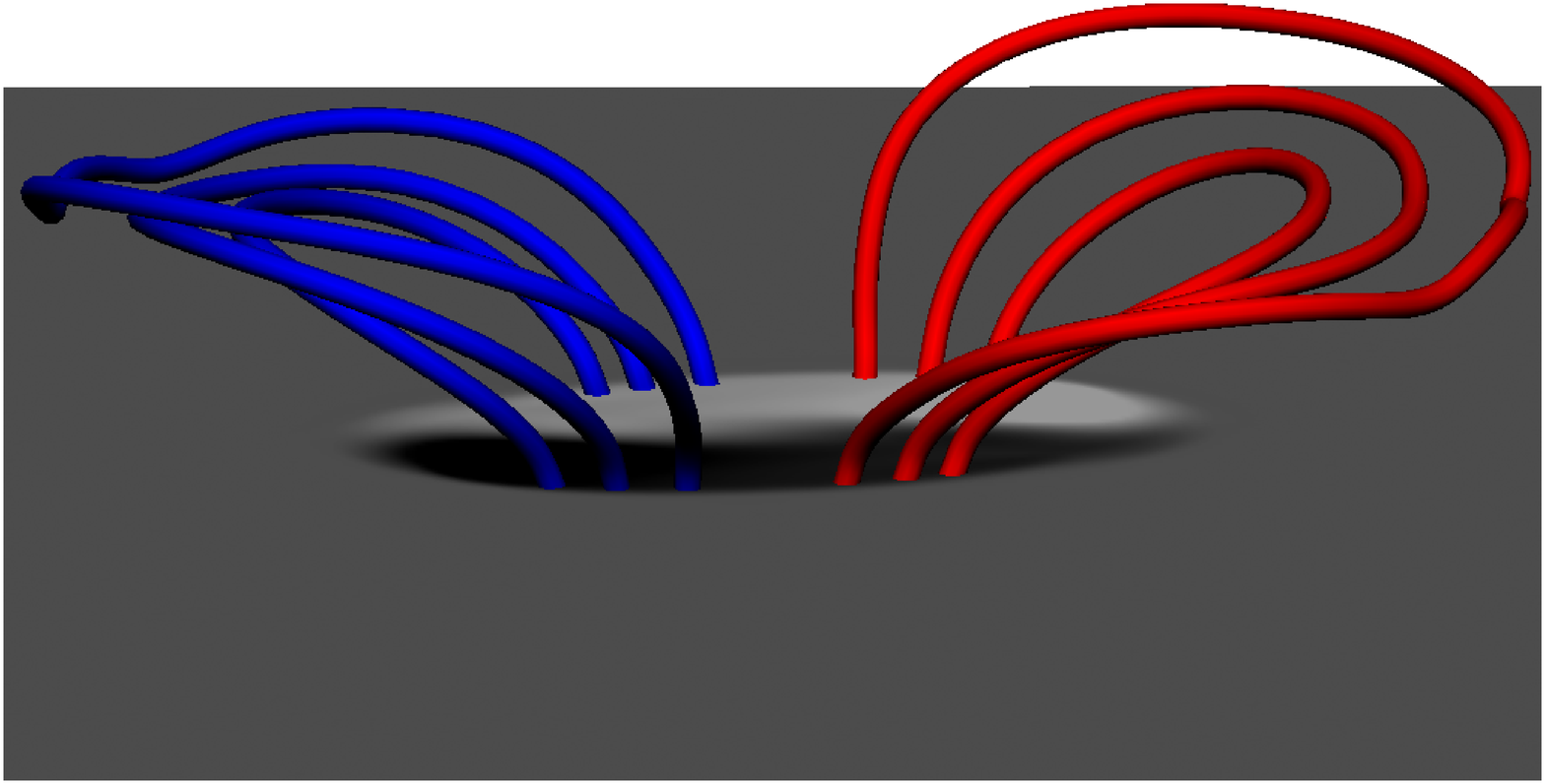}{0.42\textwidth}{(b)}
\end{center}
\caption{Representations of the magnetic field. (a) $\|\mathbf{B}\|$ in the $x=0$ plane of $E_1$ at $t=190$. (b) Selected magnetic field lines from $E_1$ at $t=190$. The slice shows $B_z$ at $z=0$. Red and blue colours indicate the two different overlying regions.}
\label{field}
\end{figure}

The process of flux emergence has been described at length in previous work \citep{hood12,cheung14}. In short, however, as the magnetic field pushes into the atmosphere, the magnetic pressure dominates the surrounding plasma pressure and can push rapidly into the corona.  Figure \ref{field} (a) displays a slice of the magnitude of the magnetic field strength in the $x=0$ plane from $E_1$ at $t=190$.

Above the photosphere $(z=0)$ there is a `magnetic bubble' that has expanded into the atmosphere. The bubble clearly expands over the footpoints (sunspots) of the emerging region. Plotting field lines, as shown in Figure \ref{field} (b) in these regions, reveals more of the geometry of the magnetic field. In particular, the inclined field has both radial and azimuthal, relative to the sunspot center, directions. This geometry could be connected to observations of \cite{lim13} which show both radially and azimuthally directed penumbrae. Figure \ref{field} demonstrates that there is a clear change in the field inclination from vertical at the footpoints to near-horizontal at the antipodal regions. We shall now give a more quantitative description of the field line inclinations in the numerical experiments.
\subsection{Probability distributions}
In order to give a quantitative measure of the field inclination, we produce kernel density estimates of the angle of the field to the vertical, $\theta=\cos^{-1}(\mathbf{B}\cdot\mathbf{e}_z/\|\mathbf{B}\|)$. The kernel density estimate (KDE) procedure generalizes the notion of a histogram \citep[e.g.][]{lindsay07}. If $\Theta_1,\dots,\Theta_n$ is a sample of $n$ observations with true density $f(\theta)$, the kernel density estimate of $f(\theta)$ is
\begin{equation}\label{kernel1}
\hat{f}(\theta)=\frac{1}{nh}\sum_{i=1}^nK\left(\frac{\theta-\Theta_i}{h}\right),
\end{equation} 
where the kernel $K(\theta)$ is non-negative and satisfies
\begin{equation}\label{kernel2}
\int^{\infty}_{-\infty}K(\theta)\,{\rm d}\theta = 1.
\end{equation}
Clearly, $\hat{f}(\theta)$ is a non-negative function that integrates to one. In equation (\ref{kernel1}) the parameter $h$ is called the bandwidth of the estimator. In order to calculate KDEs we require a particular form for the kernel function. Taking $K(\theta)$ to be the probability density function for the normal distribution with zero mean and unit variance, the KDE is
\begin{equation}\label{kernel3}
\hat{f}(\theta)=\frac{1}{nh\sqrt{2\pi}}\sum_{i=1}^n\exp\left[-\frac{1}{2}\left(\frac{\theta-\Theta_i}{h}\right)^2\right].
\end{equation} 
Following \cite{silverman86}, we choose a bandwidth that is suitable for unimodal distributions and has the form $h=1.06\sigma n^{-0.2}$ with variance $\sigma$. In this Letter, $n$ will represent the number of grid points where $\theta$ is calculated.

 In order to investigate the magnetic field inclination, we must select different regions for producing the KDEs. In each of the three experiments, we consider two regions. The first is the \emph{overlying canopy region}, OCR, which includes the AC and represents where the highly inclined field (for penumbra formation) collects. The second is the CER. For the OCR, we choose a region bounded at one side by the edge of the footpoint (near-vertical field) in the $x=0$ plane and enclosed within the photosphere/chromosphere region. In $E_1$, for example, this region is given by $(x,y,z)\in[-110,110]\times[-110,-24]\times[0,10]$ (cf. Figure \ref{field} (a)). We only consider one OCR as the other is nearly identical by symmetry. The CER is defined to be the region between the lateral boundaries of the canopy regions. For $E_1$, this is $(x,y,z)\in[-110,110]\times[-24,24]\times[0,10]$. 

\begin{figure}
\fig{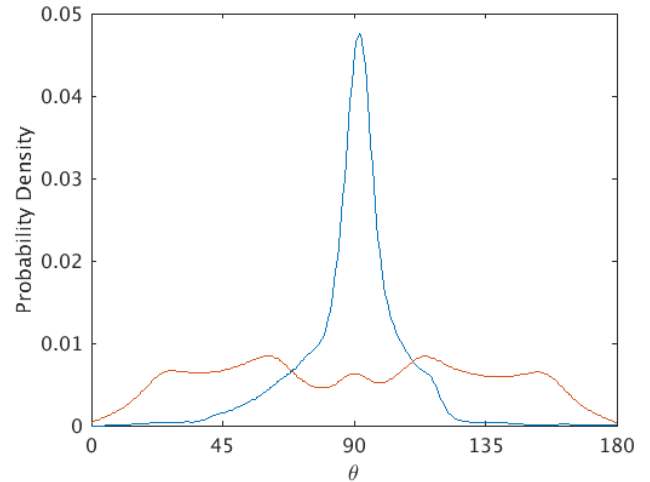}{0.5\textwidth}{(a)}\fig{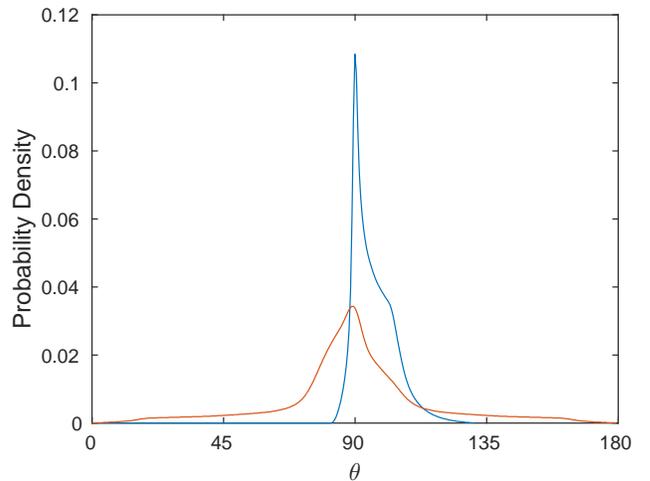}{0.5\textwidth}{(b)}
\caption{KDEs for the two regions of $E_1$. (a) shows the results from the full MHD model and (b) the potential field extrapolation. Key: OCR (blue), CER (orange).}
\label{e1}
\end{figure}

Figure \ref{e1} (a) displays $\hat{f}(\theta)$ for the two regions described above for $E_1$ at $t=190$. 
Angles close to $0^{\circ}$ or $180^{\circ}$ represent near-vertical field. Those close to $90^{\circ}$ represent near-horizontal field. When $\|\mathbf{B}\|<10^{-6}$, $\theta$ is not calculated. In the OCR there is clearly a highy probability of finding near-horizontal field and a low probability of finding near-vertical field. In the CER, there is a more uniform distribution for all inclination angles.

Figure \ref{e1} (b) displays KDEs corresponding to those in Figure \ref{e1} (a) but for a potential field extrapolation instead of the full MHD model. To calculate the potential field, we use the technique described in \cite{alissandrakis81}. On the bottom boundary, the photospheric $B_z$ profile from $E_1$ at $t=190$ is used. In calculating the potential field, the size of our computational domain is slightly different compared to the MHD simulation. However, since the magnetic field decays rapidly before it reaches the boundaries in this simulation, we do not expect this change in size to have a significant effect on the results. In the OCR, there is again a strong bias towards the field being close to horizontal. In the CER, there is a greater probability of near-horizontal field than in the MHD case. However, compared to the OCR KDE, this probability is less and there is more spread in the field inclination. Qualitatively, the results of the MHD and potential models are very similar. 

The potential field extrapolation represents an emerged field with no current density or coupling to the background plasma. The fact that this model produces results that are very similar to the full MHD case suggests that the \emph{existence} of magnetic canopies is not due primarily to the complexity of the emerged field (e.g. current structure, supporting dense plasma, etc.). Rather, it is the ease with which the emerged field can expand into the field-free atmosphere in the OCRs that facilitates the formation of highly inclined magnetic field. 

\begin{figure}
\fig{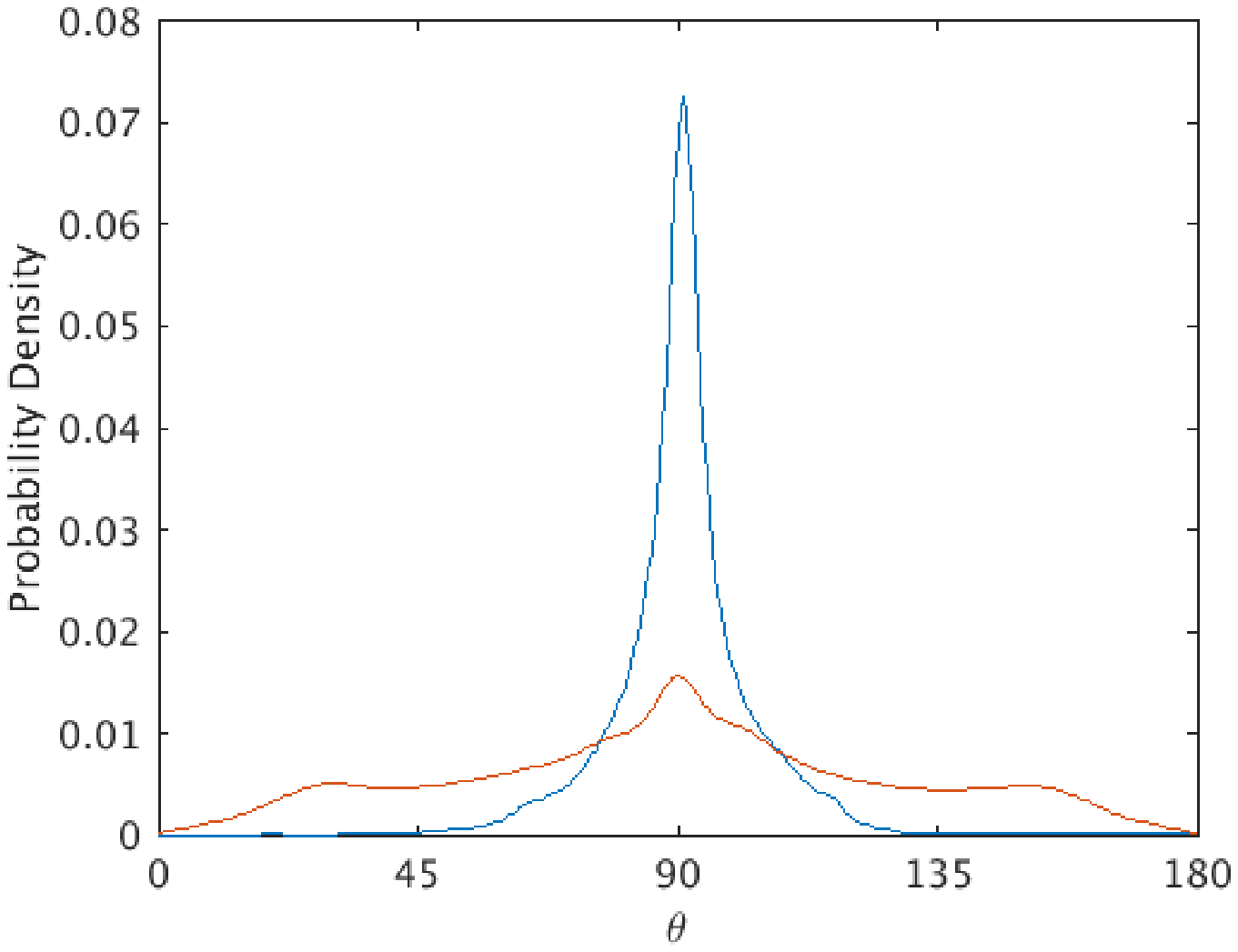}{0.5\textwidth}{(a)}\fig{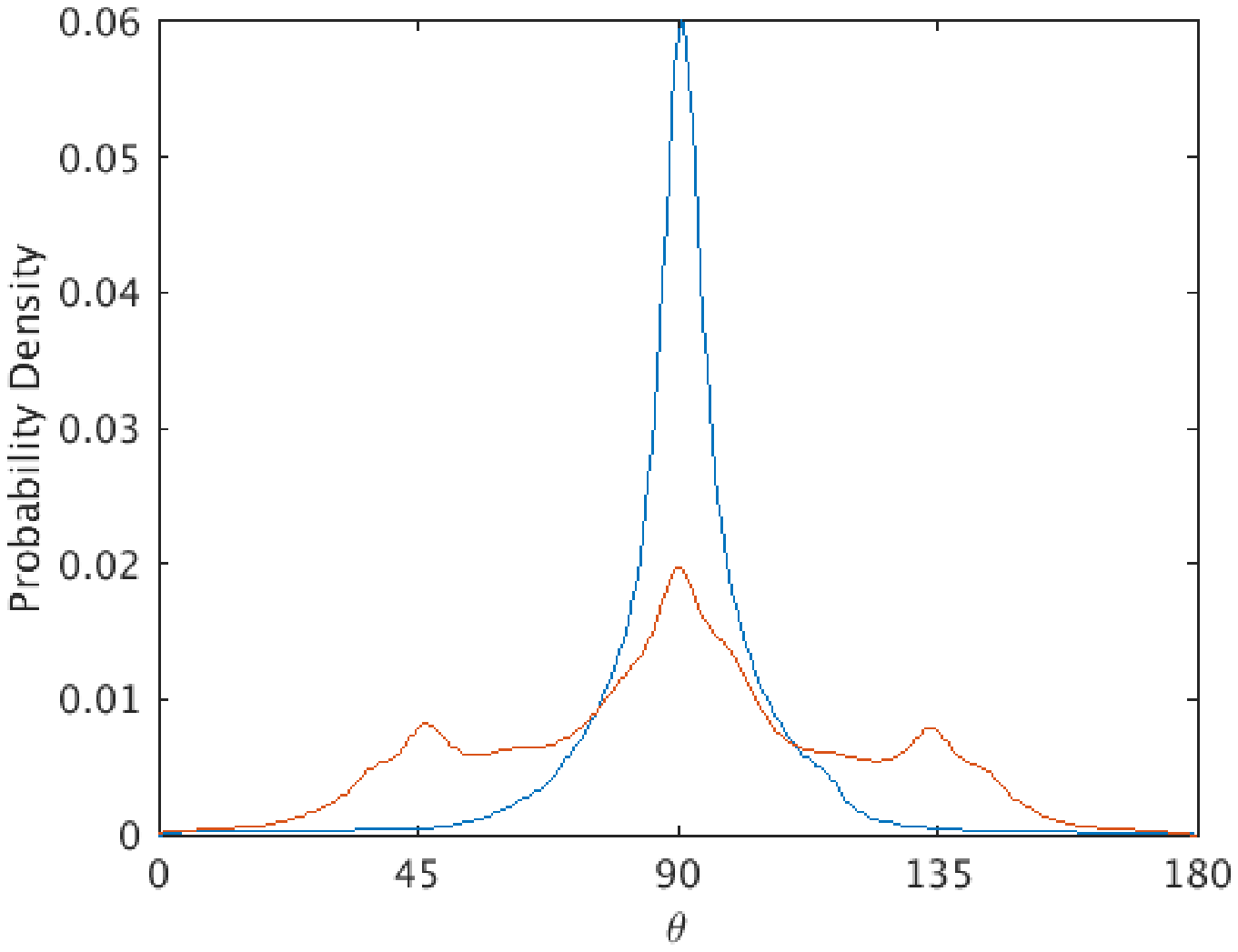}{0.5\textwidth}{(b)}
\caption{KDEs for the two regions of (a) $E_2$ at $t=190$ and (b) $E_3$ at $t=150$. Key: OCR (blue), CER (orange).}
\label{e2e3}
\end{figure}

We add weight to this result by performing two other simulations with different twist and field strength values. For these experiments we have to define different sizes for the regions as the magnetic fields expand more, within the same time period, than in $E_1$. For $E_2$, the OCR is defined by $(x,y,z)\in[-110,110]\times[-110,-52]\times[0,10]$ and the CER by $(x,y,z)\in[-110,110]\times[-52,53]\times[0,10]$. In $E_3$, the OCR is $(x,y,z)\in[-110,110]\times[-110,50]\times[0,10]$ and the CER is $(x,y,z)\in[-110,110]\times[-50,50]\times[0,10]$. These regions are selected at time $t=190$ for $E_2$ and $t=150$ for $E_3$. Since the field strength is stronger in $E_3$, its magnetic field expands faster and reaches the boundaries of the domain sooner than the others. Figure \ref{e2e3} displays the KDEs for $E_2$ and $E_3$ at the times and locations described above.
Despite some peaks appearing in the KDEs for the CERs, the general features are still very similar to results from $E_1$. The existence of distinct magnetic canopy regions in all of the numerical experiments is a robust feature. Note that we do not calculate potential field extrapolations for $E_2$ and $E_3$ as the proximity of the emerged field to the computational boundaries will bias the results.
\subsection{Canopy field strength}
In the previous section, we demonstrated that magnetic canopies can exist for different values of field strength and twist. The canopy structure is also found in a potential field extrapolation using the photospheric boundary of $E_1$. Although the existence of magnetic canopies does not appear to be sensitive to the complexity of the emerged field, the formation of penumbrae will be affected. \cite{rempel12} found that in order to produce extended penumbrae, the horizontal field (canopy) has to have a field strength that is approximately twice that of an equivalent potential field. In order to assess the effects of current density and plasma coupling on the canopy field strength, we shall present three cases from $E_1$ at $t=190$. The first case is the potential field extrapolation discussed in the previous section. A potential field is one with no current or coupling to the background plasma and represents the extreme case of field relaxation. The second case is the full MHD model, where the magnetic field has a current density structure and also supports dense plasma, carried upwards from the photosphere during emergence. The third case represents a scenario somewhere between the first two cases. In order produce a field that is twisted but does not support any dense plasma, we re-run $E_1$ with the modified mass conservation equation,
\begin{equation}\label{newton}
{\dot{\rho}} = -\rho\nabla\cdot\mathbf{u} -\frac{\rho-\rho_0}{\tau}.
\end{equation}
In equation (\ref{newton}) we have added a relaxation term to drive the density to $\rho_0$, its value at $t=0$. The rate of relaxation is governed by $\tau$. In this Letter, we set $\tau=0.5$ throughout the domain. Doing so allows for the density to relax rapidly to its initial condition on a time scale much faster than that of the formation of magnetic canopies. The result of running simulation $E_1$ with equation (\ref{newton}) rather than equation (\ref{mass_con}) is that the emerged field supports no dense plasma carried upwards from the photosphere. That is, draining is completely efficient and the density profile in the atmosphere at $t=190$ is the same as it was at $t=0$.

\begin{figure}
\plotone{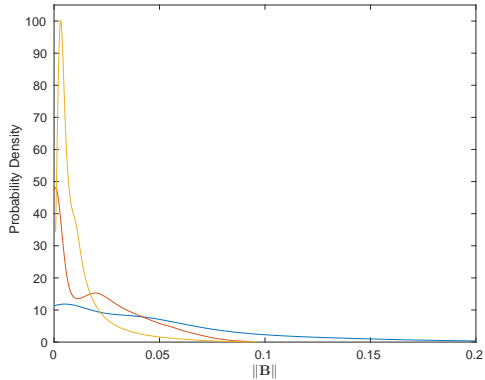}
\caption{KDEs of $\|\mathbf{B}\|$ in an OCR. Key: potential (yellow), full MHD(blue), modified density (orange).}
\label{bmag}
\end{figure}

In order to compare the canopy field strengths for the three cases listed above, we plot KDEs of $\|\mathbf{B}\|$ in an OCR. For the first and third cases, the dimensions of the OCR are as stated previously. For the third case, the field expands more by $t=190$ compared with the other cases and the limits in the $y$-direction are now slightly modified to $[-110,-29]$. Figure \ref{bmag} displays the distributions of magnetic field strengths in an OCR for each of the three cases.
From Figure \ref{bmag}, the typical field strength values of the potential case (yellow) are the weakest out of the three cases. Its KDE decays before $\|\mathbf{B}\|=0.1$. The modified density case (orange) also has a KDE that decays before $\|\mathbf{B}\|=0.1$ and is concentrated at weak field strengths. However, in the modified density KDE, there is a greater probability of finding higher field strengths of $\|\mathbf{B}\|\approx 0.05$ compared to the potential case. The full MHD case (blue) KDE has a much larger spread in field strength values and extends to values much larger than the other cases.

The above KDEs profiles can be interpreted in terms of the complexity of the emerged field. The potential case has no current and does not support dense plasma. It represents a minimum-energy solution and so has the weakest field strength values. The modified density case mangetic field has twist (non-zero current) but does not support dense plasma. The twist in the field allows for greater field strengths compared to the potential case. Finally, the full MHD case has an emerged field that is both twisted and supports dense plasma. The effect of the dense plasma on the canopies is to compress the field and produce stronger field strengths. The values found in the full MHD case can be an order of magnitude greater than those in the potential case. Hence, the combination of twist and plasma coupling in the emerged field can produce field strengths required for the formation of extended penumbrae \citep{rempel12}.

\section{Discussion}
In this Letter we have presented simulations of flux emergence and have demonstrated, through visualizations and KDEs of the field inclination, that they produce near-horizontal magnetic canopies at the antipodal curves of the footpoints. Several observational studies \citep[e.g.][]{shimizu12, romano13} suggest that an overlying magnetic canopy is required to produce penumbrae. We show that the existence of such magnetic canopies is not sensitive to the complexity of the emerged field. The field strength of the canopies, which will influence the development of penumbrae, does, however, depend on the complexity of the emerged field. By considering three magnetic field models - potential, twisted but not supporting dense plasma, twisted and supporting dense plasma - we demonstrate that the inclusion of twist and plasma coupling can produce canopy field strengths greater than double the equivalent potential values. This means that current and plasma coupling in the emerged field can produce canopies that can, in turn, lead to the formation of extended penumbrae \citep{rempel12}.

Although the simulations we present here are highly idealized and cannot produce the fine-scale structure of sunspots, they have the advantage of being able to isolate particular physical processes whilst still being able to describe the large-scale features of flux emergence. One simplification that was made was to consider regions smaller than a typical active region. This decision was made in order to adequately resolve the photosphere/chromosphere region. It was shown that increasing the field strength does not alter the formation of magnetic canopies. Indeed, the canopies grow more rapidly due to the faster expansion of the stronger emerging field \citep{murray06}. 

We expect our results to survive the inclusion of extra physics in the model. The inclusion of convection \citep[e.g.][]{rempel14} will make emergence within the CER more complex. However, if the field is strong enough, convection should not prevent its expansion into the atmosphere and, hence, the formation of canopies.

We also note here that our full MHD simulations can over-estimate the amount of dense plasma carried into the atmosphere \citep[e.g.][]{arber07}. However, our modified density model shows that canopies still form even if draining is completely efficient.

\acknowledgements

We thank the Referee for many valuable comments. We acknowledge a Carnegie Trust Research Incentive Grant (Ref: 70323) and SOLARNET (http://www.solarnet-east.eu),  funded  by the European  Commision’s  FP7  Capacities  Programme  under  the  grant  agreement No. 312495. Computational resources were provided by the EPSRC funded ARCHIE-WeSt High Performance Computer (www.archie-west.ac.uk), EPSRC grant no. EP/K000586/1. This work was also supported by the Italian MIUR-PRIN grant 2012P2HRCR on \textit{The active Sun and its effects on Space and Earth climate} and by Space WEather Italian COmmunity (SWICO) Research Program. The SDO/HMI data are courtesy of NASA/SDO and the HMI science team.


\begin{thebibliography}{}
\bibitem[Alissandrakis(1981)]{alissandrakis81} Alissandrakis, C.~E. \ 1981 \aap, 100, 197

\bibitem[Arber et al.(2001)]{arber01} Arber, T.~D., Longbottom, A.~W., Gerrard C.~L., et al. \ 2001, J. Comput. Phys., 171, 151

\bibitem[Arber et al.(2007)]{arber07} Arber, T.~D., Haynes, M. \& Leake, J.~E.  \ 2007, \apj, 666, 541

\bibitem[Borrero \& Ichimoto(2011)]{borrero11} Borrero, J.~M. \& Ichimoto, K. \ 2011, Living Rev. Sol. Phys., 8, 4

\bibitem[Cheung \& Isobe(2014)]{cheung14} Cheung, M.~C.~M. \& Isobe, H. \ 2014, Living Rev. Sol. Phys., 11, 3


\bibitem[Galsgaard et al.(2005)]{galsgaard05} Galsgaard, K., Moreno-Insertis, F., Archontis, V., et al. \ 2005, \apjl, 618, L153

\bibitem[Guglielmino et al.(2010)]{guglielmino10} Guglielmino, S.~L., Bellot Rubio, L.~R., Zuccarello, F., et al. \ 2010, \apj, 724, 1083

\bibitem[Hood et al.(2012)]{hood12} Hood, A~W., Archontis, V. \& MacTaggart, D. \ 2012, \solphys, 278, 3


\bibitem[Lim et al.(2013)]{lim13} Lim, E.-K., Yurchyshyn, V., Goode, P., et al. \ 2013, \apjl, 769, L18

\bibitem[Lindsay et al.(2007)]{lindsay07} Lindsay, K.~A., Maxwell, D.~J., Rosenberg, J.~A. et al. \ 2007, Math. Biosci., 205, 271 

\bibitem[MacTaggart et al.(2015)]{dmac15} MacTaggart, D., Guglielmino, S.~L., Haynes, A.~L., et al. \ 2015, \aap, 556, A40


\bibitem[Murabito et al.(2016)]{murabito16} Murabito, M., Romano, P., Guglielmino, S.~L., et al. \ 2016, \apj, 825, 75

\bibitem[Murray et al.(2006)]{murray06} Murray, M.~J., Hood, A.~W., Moreno-Insertis, F. et al. \ 2006, \aap, 460, 909

\bibitem[Rempel \& Schlichenmaier(2011)]{rempel11} Rempel, M. \& Schlichenmaier, R. \ 2011, Living Rev. Sol. Phys., 8, 3

\bibitem[Rempel(2012)]{rempel12} Rempel, M. \ 2012, \apj, 750, 62

\bibitem[Rempel \& Cheung(2014)]{rempel14} Rempel, M. \& Cheung, M.~C.~M. \ 2014, \apj, 785, 90

\bibitem[Rezaei et al.(2012)]{rezaei12} Rezaei, R., Bello Gonz\'{a}lez \&  Schlichenmaier, R. \ 2012, \aap, 536, A19

\bibitem[Romano et al.(2013)]{romano13} Romano, P., Frasca, D., Guglielmino, S.~L., et al. \ 2013, \apj, 77, L3

\bibitem[Schlichenmaier et al.(2010)]{schlichenmaier10} Schlichenmaier, R., Rezaei, R., Bello Gonz\'{a}lez et al. \ 2010, \aap, 512, L1 

\bibitem[Shimizu et al.(2012)]{shimizu12} Shimizu, T., Ichimoto, K. \& Suematsu, Y. \ 2012, \apjl, 747, L18

\bibitem[Silverman(1986)]{silverman86} Silverman, B.~W. \ 1986, Density estimation for statistics and data analysis, Monographs on Statistics and Applied Probability, London: Chapman \& Hall

\bibitem[Thomas \& Weiss(2008)]{thomas08} Thomas, J.~H. \& Weiss, N.~O. \ 2008, Sunspots and Starspots, Cambridge University Press


\end{thebibliography}
\end{document}